\documentclass{ifacconf}

\usepackage{graphicx}      % include this line if your document contains figures
\usepackage{natbib}        % required for bibliography

\usepackage{graphicx}
\usepackage{textcomp}
\usepackage{xcolor}
\usepackage{mathtools}
\usepackage{xfrac}
\usepackage{comment}
\usepackage{pgfplots}
\pgfplotsset{compat=1.18}
\usepackage{pgfmath}
\usepackage{pgfplotstable}
\usetikzlibrary{calc,decorations.markings,intersections}
\usepgfplotslibrary{fillbetween,patchplots}
\usepackage{amsmath,amssymb,amsfonts}
\newcommand\scalemath[2]{\scalebox{#1}{\mbox{\ensuremath{\displaystyle #2}}}}

\usepackage{algorithm}
\usepackage{algpseudocode}
\algrenewcommand\alglinenumber[1]{\scriptsize\textcolor{gray}{#1:}}
\usepackage{booktabs}
\usepackage{url}

\graphicspath{{..}}

\newcommand{\R}{\mathbb{R}}
\newcommand{\U}{\mathbb{U}}
\newcommand{\Y}{\mathbb{Y}}

\newcommand{\tr}{\intercal}

\definecolor{sampathcolor}{HTML}{ff6f69}

\newcommand{\norm}[1]{\left\lVert#1\right\rVert}

\DeclareFontFamily{U}{boondoxuprscr}{\skewchar \font =45}
\DeclareFontShape{U}{boondoxuprscr}{m}{n}{
    <-> BOONDOXUprScr-Regular}{}
\DeclareFontShape{U}{boondoxuprscr}{b}{n}{
    <-> BOONDOXUprScr-Bold}{}

\newtheorem{remark}{Remark}
\newtheorem{assumption}{Assumption}

\usepackage{multirow}

\begin{document}
\begin{frontmatter}

\title{A Regularization and Active Learning Method for Identification of Quasi Linear Parameter Varying Systems} 

\thanks[footnoteinfo]{This work was funded by the European Union (ERC Advanced Research Grant COMPACT, No. 101141351). Views and opinions expressed are however those of the authors only and do not necessarily reflect those of the European Union or the European Research Council. Neither the European Union nor the granting authority can be held responsible for them.}

\author[First]{Sampath Kumar Mulagaleti} 
\author[First]{Alberto Bemporad} 

\address[First]{ IMT School for Advanced Studies Lucca, Italy}

\begin{abstract}              
This paper proposes an active learning method for designing experiments to identify quasi–Linear Parameter-Varying (qLPV) models. Since informative experiments are costly, input signals must be selected to maximize information content based on the currently available model. To improve the extrapolation properties of the identified model, we introduce a manifold-regularization strategy that enforces smooth variations in the qLPV dynamics, promoting Linear Time-Varying (LTV) behavior. Using this regularized structure, we propose a new active learning criterion based on path integrals of an inverse-distance variance measure and derive an efficient approximation exploiting the LTV smoothness. Numerical examples show that the proposed regularization enhances qLPV extrapolation and that the resulting active learning scheme accelerates the identification process.
\end{abstract}


\end{frontmatter}
%===============================================================================
\section{Introduction}
The selection of a dynamical model for model-based control inherently involves a trade-off between accuracy, which is achievable with detailed nonlinear representations of the plant, and simplicity, which in turn makes the control design simpler, especially when online dynamic optimization is employed. In this regard, quasi Linear Parameter Varying (qLPV) \citep{Toth2010Modeling} models have emerged as an effective choice. In qLPV systems, also known as self scheduled LPV systems, the dynamics are described by linear models that change over time as a function of the scheduling vector, whose values are generated by a nonlinear function of the model state and input. This has led to increased interest in the identification of such models \citep{Verhoek2022}.

System identification performance is heavily influenced by the availability of informative experimental data. Since experimentation is often costly or time-consuming, it is crucial to apply input signals that sufficiently excite the plant so that a reliable model can be obtained with as few experiments as possible. A common approach is optimal experiment design \citep{goodwin1977dynamic}. In this paradigm, the currently identified model is used to compute an input sequence that is then applied to the plant; the resulting output is collected and appended to the dataset, after which the model is re-identified. The underlying idea, which is well known in the machine learning literature as active learning \citep{settles2009active}, is that carefully optimized inputs yield data that improve subsequent model estimates.

For linear system identification, several methods have been developed based on the Fisher information matrix \citep{Gevers2011,wagenmaker20a}, which aim to compute inputs that minimize the parameter covariance. Unfortunately, these methods are not straightforwardly applicable to nonlinear system identification. While there have been attempts to extend the theoretical results to the nonlinear case \citep{wagenmaker2023optimal,mania2022active}, they rely on restrictive model classes.
%An active learning method for an externally-scheduled LPV system was introduced in \citep{Chin2020}.

A fundamental challenge in active learning arises from its goal of training models that reliably extrapolate to unseen scenarios while limiting the amount of training data collected. However, models trained on sparse datasets typically struggle to predict outside the training domain, yet at the same time, designing high-quality input sequences still requires a model that can extrapolate reliably. In the linear case, this issue has been addressed in \citep{Rojas2007}, which extends the Fisher approach to consider a min–max optimal input design over a set of parameters. In the general case, regularization can be employed to improve extrapolation capabilities. As noted in \citep{settles2009active}, semi-supervised learning, which leverages unlabeled data, can provide an effective approach to derive such regularization functions. Of particular relevance to this paper is manifold regularization \citep{Belkin2006}, which exploits unlabeled samples to ensure that the learned function varies smoothly along the data manifold. This idea has been effectively used for the development of training and active learning strategies \citep{Zhou2003,He2010,zhang2016manifold}, and specifically for system identification \citep{Ohlsson1008,Formentin2019}. A modern approach leveraging synthetic input–output data to achieve similar ends was presented in \citep{Piga2024}. These approaches, however, were not developed for system identification with qLPV models.

The main contributions of this paper are the following:
\begin{enumerate}
\item [($i$)] We develop a regularization strategy based on manifold regularization to improve the extrapolation performance of qLPV models. The idea is to make the qLPV system behave as a smoothly varying Linear Time-Varying (LTV) system, that is, with linear matrices that vary smoothly with the input sequence.
\item [($ii$)] We introduce an active learning criterion, inspired by the inverse-distance weighted variance formulation in \citep{bemporad2023}, based on path integrals along the graph of the qLPV system, together with an approach to simplify the evaluation of the criterion by exploiting the LTV regularization.
\end{enumerate}
Through numerical examples, we show that the proposed regularization helps identify qLPV models with improved extrapolation quality. Furthermore, we demonstrate that the proposed active learning criterion helps select input sequences that accelerate the identification procedure.

\textit{Notation:} Given vectors $x_i \in \R^n$ for $i \in \{1,\ldots,N\}$, we denote by $(x_1,\ldots,x_N) \in \R^{Nn}$ the vector obtained by vertical stacking. The symbol $\oplus$ indicates the Minkowski sum operator, and $\mathbb{B}^n_{\infty}$ represents the $\infty$-norm ball in $\R^n$. Given sets $\mathbb{S}_i \in \R^{n}$ for $i \in \{1,\ldots,N\}$, the symbol $\Pi_{i=1}^N \mathbb{S}_i$ denotes their Cartesian product in $\R^{Nn}$.
\section{Background}
Suppose we have a dynamical system of the form
\begin{align}
\label{eq:underlying}
\mathbf{z}^+ = \mathbf{f}(\mathbf{z},u), && y=\mathbf{g}(\mathbf{z})
\end{align}
where the state $\mathbf{z}$, the transition function $\mathbf{f}$, and the output function $\mathbf{g}$ are unknown, and the input $u \in \R^{n_u}$ and output $y \in \R^{n_y}$ can be measured. Furthermore, the input and output are subject to pointwise-in-time constraints $u\in\U$ and $y\in\Y$ respectively. Our goal is to identify a dynamical model of this system. In particular, we want to identify a qLPV model 
\begin{align}
    \label{eq:qLPV}
    x^+=A(p(x,u))x+B(p(x,u))u, && y=Cx
\end{align}
to predict the dynamics of \eqref{eq:underlying} over a fixed horizon length $T \in \mathbb{N}$, where we parameterize $A(p)$ and $B(p)$ as
\begin{align}
    \label{eq:qLPV_sum}
    A(p):=\sum_{i=1}^{n_p} p_i A_i, && B(p):=\sum_{i=1}^{n_p} p_iB_i,
\end{align}
with each $A_i \in \R^{n_x \times n_x}$, $B_i \in \R^{n_x \times n_u}$ and $C \in \R^{n_y \times n_x}$. Following \citep{mulagaleti2024a,mulagaleti2025}, we parameterize the scheduling function as
\begin{align}
    \label{eq:scheduling}
    p(x,u) := \sigma(h(x,u)),
\end{align}
where $h(\cdot,\cdot):\R^{n_x} \times \R^{n_u} \to \R^{n_p}$ is a feedforward neural network (FNN), and $\sigma(\cdot):\R^{n_p} \to \R^{n_p}$ 
is the $\mathrm{softmax}$ operator. This parameterization ensures that
\begin{align*}
    p(x,u) \in \Delta:=\left\{p \geq 0 \middle| \sum_{i=1}^{n_p}p_i=1\right\}, \ \  \forall (x,u) \in \R^{n_x} \times \R^{n_u},
\end{align*}
where $\Delta$ is the unit simplex.

\textbf{\textit{Identification setting:}} Let $U=(u_0,\ldots,u_{T-1}) \in \R^{Tn_u}$ and $Y=(y_0,\ldots,y_{T-1}) \in \R^{Tn_y}$ and assume that the underlying plant to be identified is always initialized at the origin, i.e., $\mathbf{z}_0=0$ (although we will present a preliminary approach to extend the setting to nonzero initial states in Section \ref{sec:cascaded}, this a subject of future research.) Let $\mathbf{F}:\R^{Tn_u} \to \R^{Tn_y}$ be the function obtained by propagating the equations in \eqref{eq:underlying} and define the set
\begin{align*}
    \mathbf{U}:=\left\{ U \in \U_T \middle| \mathbf{F}(U) \in \Y_T\right\},
\end{align*}
where $\U_T := \Pi_{t=1}^T \U \subseteq \R^{Tn_u}$ and $\Y_T := \Pi_{t=1}^T \Y\subseteq \R^{Tn_y}$. The set $\mathbf{U}$ captures all input trajectories resulting in output trajectories satisfying the constraints for $T$ time steps. Denoting the parameters of model \eqref{eq:qLPV} as $\theta \in \R^{n_\theta}$ and assuming that \eqref{eq:qLPV} is always initialized with $x_0=0$ as the underlying plant, we construct the function $F(\cdot|\theta):\R^{Tn_u} \to \R^{Tn_y}$ by propagating \eqref{eq:qLPV}. By defining the cost function
\begin{align*}
    \ell(\theta):= \int_{U \in \mathbf{U}} \left[\|\mathbf{F}(U)-F(U|\theta)\|_2^2\right] dU,
\end{align*}
we aim to solve the optimization problem
\begin{align}
\label{eq:ideal_problem}
    \theta_*:=\arg\min_{\theta} \ \ell(\theta).
\end{align}
\begin{remark}
Problem \eqref{eq:ideal_problem}, ideally, must be formulated with the constraint $f(U|\theta) \in \Y_T$ for all $U \in \mathbf{U}$ to ensure that the model satisfies the plant constraints. Incorporating such constraints is a subject of future research. 
\end{remark}
To solve Problem \eqref{eq:ideal_problem} empirically, we assume to have access to a dataset
\begin{align*}
    \mathcal{D}_{N}:=\left\{(U_i,Y_i) \middle| U_i \in \mathbf{U}, Y_i=\mathbf{F}(U_i),i \in \{1,\ldots,N_{\mathrm{d}}\}\right\}
\end{align*}
of $N_{\mathrm{d}}$ input-output trajectories recorded from the plant, using which we formulate the problem
\begin{align}
\label{eq:sysID_problem}
    \theta_N:=\arg\min_{\theta} \frac{1}{N_{\mathrm{d}}} \sum_{i=1}^{N_{\mathrm{d}}}\|Y_i-F(U_i|\theta)\|_2^2 + r(\theta),
\end{align}
where $r(\theta)$ is a regularization on the model parameters that we develop in the sequel. 

\subsection{Active learning}
This paper addresses the development of active learning (AL) algorithms, a class of methods that iteratively compute informative inputs to improve model accuracy. At each iteration, given the current model $Y=F(U|\theta_N)$, a new input $U_{\mathrm{a}}$ is determined and applied to the plant, yielding a new output measurement $Y_{\mathrm{a}}$. The input is selected by optimizing an acquisition function $a(U|\theta_N,\mathcal{D}_N)$:
\begin{align}
    \label{eq:AL_problem}
    U_{\mathrm{a}}:=\arg\max_{U \in \mathbf{U}} \ a(U|\theta_N,\mathcal{D}_N).
\end{align}
Let $\theta_{N+1}$ denote the optimizer of Problem \eqref{eq:sysID_problem} and let $\mathcal{D}_{N+1}=\mathcal{D}_N \cup \{(U_{\mathrm{a}},\mathbf{F}(U_{\mathrm{a}}))\}$. The acquisition function $a$ is designed to maximize the expected reduction in the loss function, i.e., to minimize $\ell(\theta_{N+1})-\ell(\theta_N)$. 
We use a model-based active learning method, which aims to compute an input sequence that explores the combined input-output space as much as possible. A common heuristic is to select a new input–output pair $(U_{\mathrm{a}},Y_{\mathrm{a}})$ that lies \textit{far} from existing trajectories contained in $\mathcal{D}_N$, using the model prediction $F(U_{\mathrm{a}}|\theta_N)$ as a surrogate for $Y_{\mathrm{a}}$. This introduces a fundamental trade-off between \textit{informativity} and \textit{diversity}: If $U_{\mathrm{a}}$ is chosen too far from the previous samples, the model prediction may be unreliable. Then, no conclusion can be drawn whether this prediction is sufficiently different from the plant output; conversely, if it is too close, the new data provide little additional information. To address this trade-off, we first introduce a regularization mechanism that enlarges the region in which the model predictions can be considered reliable. Then, we propose a novel distance metric for the acquisition function based on approximating the path integral along the model manifold. For both these contributions, we exploit the qLPV nature of our model parameterization.

\section{LTV-modeling based regularization}
\label{sec:LTV_reg}
Let $\mathcal{D}_N$ be a given dataset of input–output measurements from which we have identified a model. As we use the latter as a surrogate of the underlying plant, we work under the following assumption.
\begin{assumption}
\label{ass:LTV_assumption}
    Plant \eqref{eq:underlying} is a smoothly varying LTV system, i.e., for input trajectories $U^1$ and $U^2$ that are close to each other, the corresponding outputs $\mathbf{F}(U^1)$ and $\mathbf{F}(U^2)$ are generated by LTV systems with sequences of system matrices that are close to each other.
\end{assumption}
Assumption \ref{ass:LTV_assumption} is called the \textit{manifold assumption}, and is regularly employed in manifold regularization literature \citep{Formentin2019}. We remark that a wide range of Lipschitz continuous systems satisfy this assumption, and is the basis of many LPV control schemes \citep{ALCALA2019}. Based on this assumption,
we design a regularization function $r(\theta)$ to employ in Problem \eqref{eq:sysID_problem}, with the aim of avoiding large changes of the scheduling-variable sequence
%enhancing the \textit{reliability} of the model predictions 
in neighborhoods of the input trajectories contained in the dataset. 
%We adopt the following informal notion of reliability $-$
For a fixed scheduling-variable sequence, the qLPV system in \eqref{eq:qLPV} reduces to an LTV system. Thus, by regularizing the scheduling map, we obtain a qLPV model with locally consistent LTV dynamics. We define the set of input sequences in the neighborhood of the dataset as
\begin{align}
\label{eq:neighborhood_U}
\bar{\mathbb{U}}_N := \bigcup_{(U,Y) \in \mathcal{D}_N}\left\{\{U\} \oplus \epsilon_{\mathrm{u}} \mathcal{B}_{\infty}^{Tn_u}\} \cap \mathbb{U}_T\right\},
\end{align}
where $\epsilon_{\mathrm{u}} \ge 0$ is a user-specified neighborhood radius. 
We say that the identified qLPV system is \textit{gradually varying} in $\bar{\mathbb{U}}_N$ if similar input sequences within this set yield similar scheduling-variable sequences, and, therefore, model matrices.
A straightforward way to enforce such smooth changes is to employ an $\ell_2$ regularization of the model parameters, i.e., $r(\theta) = \kappa \|\theta\|_2^2$. For sufficiently large $\kappa$, this choice drives $\theta \approx 0$, thereby promoting flatness. However, this comes at the cost of introducing a significant model bias. Our goal is therefore to design a regularization function that ensures smoothness in $\bar{\mathbb{U}}_N$ while mitigating excessive bias. To this end, we exploit the specific qLPV structure of the model parameterization. 

We first observe that since the initial state $x_0=0$, the function $F(U)$ is obtained by propagating the dynamical system in \eqref{eq:qLPV} as
\begin{align*}
    F(U)=\scalemath{1.}{\begin{bmatrix*}[l] 
                 0 \\
                 CB(p_0) \\
                 CA(p_1)B(p_0) & CB(p_1) \\ 
                 \vdots & \vdots & \ddots \\
                 C\underset{t=T-1}{\overset{1}{\prod}}A(p_t)B(p_0) & \hdots & \hdots & CB(p_{T-1})
        \end{bmatrix*}}U,
\end{align*}
where $p_t=p(x_t,u_t)$ is the value of the scheduling function obtained by propagating \eqref{eq:qLPV}, with $x_t$ denoting the state at time $t$. Then, denoting the function
\begin{align*}
    P(U|\theta):=(p(0,u_0),\ldots,p(x_{T-1},u_{T-1})),
\end{align*}
we note that $P(\cdot|\theta):\R^{Tn_u} \to \Delta_T:=\Pi_{t=1}^T\Delta \subset \R^{Tn_p}$ because of \eqref{eq:scheduling}. This implies that we can write
\begin{align}
    \label{eq:main_observation}
F(U|\theta) = G(P(U|\theta))U,
\end{align}
where $G(\cdot|\theta):\Delta_T \to \R^{Tn_y \times Tn_u}$ is a matrix-valued function. Equation \eqref{eq:main_observation} highlights that, for a fixed scheduling trajectory $P(U|\theta)$, the qLPV system behaves as an LTV system. To analyze how $F(U|\theta)$ varies with the input sequence, we make the following assumption.
\begin{assumption}
    \label{ass:smooth}
    The function $P(U|\theta)$ is differentiable in $U$.
\end{assumption}
Let us denote 
\begin{align}
\label{eq:Lambda_defn}
\Lambda(U|\theta):=\nabla_U \left[G(P(U|\theta))U\right] \in \R^{Tn_y \times Tn_u}.
\end{align}
Applying the chain rule, we obtain
\begin{align}
\label{eq:key_eqn}
\hspace{-8pt}
    \Lambda(U|\theta) = G(P(U|\theta))+\sum_{i=1}^{Tn_p} \frac{\partial G(P(U|\theta))}{\partial P_i} \nabla_U P_i(U|\theta),
\end{align}
with $\partial G(P(U))/\partial P_i \in \R^{Tn_y \times 1}$ and $\nabla_U P_i(U) \in \R^{1 \times Tn_u}$. Recalling our aim, we seek to ensure that $\Lambda(U|\theta)$ remains small for all input trajectories $U \in \bar{\mathbb{U}}_N$ since small values of $\Lambda(U|\theta)$ imply that the input-output mapping $F(U|\theta)$ varies smoothly with $U$, thereby guaranteeing that similar input sequences yield similar LTV systems. While a standard $\ell_2$-regularization on model parameters reduces the magnitudes of $G(P(U|\theta))$ and $\partial G(P(U|\theta))/\partial P_i$, its influence on the terms $\nabla_U P_i(U|\theta)$ is far less direct. To explicitly reduce the latter contributions, i.e., minimize how the scheduling variables change with the input sequence, we propose the following strategy.
We sample a large pool of input trajectories from $\bar{\mathbb{U}}_N$ and build the set
\begin{align*}
    \mathcal{U}_N := \left\{U_k \in \bar{\mathbb{U}}_N \middle| k=\{1,\ldots,N_{\mathrm{r}}\}\right\}.
\end{align*}
Then, we can define the regularization function as
\begin{align}
\hspace{-5pt}
    \label{eq:regularization_1}
    r(\theta)=\kappa_1 \|\theta\|_2^2 + \frac{\kappa_2}{N_{\mathrm{r}}} \scalemath{0.9}{\sum_{k=1}^{N_{\mathrm{r}}} \sum_{i=1}^{Tn_p}} \nabla_U P_i(U_k|\theta) \nabla_U P_i(U_k|\theta)^{\tr}.
\end{align}
 While \eqref{eq:regularization_1} is a viable regularization function, it can be computationally very expensive to evaluate and optimize. An alternative is to employ a regularization inspired by semi-supervised learning and manifold regularization techniques, see \citep{Belkin2006}. The smoothness of $P(U|\theta)$ with respect to $U$ can be promoted by minimizing
\begin{align}
\label{eq:m_formulation}
    m(\theta):=\frac{1}{N_{\mathrm{r}}}\sum_{k=1}^{N_{\mathrm{r}}-1}\sum_{l=k+1}^{N_{\mathrm{r}}} \frac{ \|P(U_k|\theta)-P(U_l|\theta)\|_2^2}{\|U_k-U_l\|_2^2}.
\end{align}
Essentially, $m(\theta)$ is formulated in such a way that points $U_k$ and $U_l$ that are close to each other will result in similar scheduling variable sequences, since their corresponding weight $1/\|U_k-U_l\|_2^2$ would be higher.
This choice yields the regularization
\begin{align}
    \label{eq:regularization_2}
    r(\theta)=\kappa_1 \|\theta\|_2^2 + \kappa_2 m(\theta).
\end{align}
 Note that \eqref{eq:m_formulation} is one way to model the smooth property; other approaches based on exploiting underlying geometry are viable alternatives. We also remark that, while we wish $\epsilon_u$ in~\eqref{eq:neighborhood_U} be large, this would result in requiring a large number of unlabeled trajectories $N_{\mathrm{r}}$ to densely cover the set.

\section{Path length approximations}
\label{sec:integral_approximation}
Consider a fixed parameter vector $\theta$ and let $F_{\theta}(U) = F(U|\theta)$ for simplicity of notation. 
%We further assume that the model perfectly represents the underlying plant, i.e., $F(U) = \mathbf{F}(U)$. This assumption will be relaxed in the next section.
As discussed previously, the acquisition function $a$ in Problem \eqref{eq:AL_problem} is designed to compute a new point as far as possible from existing samples in the joint input-output domain, i.e., the graph $S$ of $F_{\theta}$, where $F_{\theta}(U)=G(P(U))U$, defined as
\begin{align*}
S := \left\{ (U, Y) \in \R^{Tn_u} \times \R^{Tn_y} \middle| Y = G(P(U))U, U \in \mathbb{U}_T \right\}.
\end{align*}
Accordingly, the function $a$ is defined in terms of path lengths measured on this graph.

Consider two points $(U^{\mathrm{1}},Y^{\mathrm{1}}) \in S$ and $(U^{\mathrm{2}},Y^{\mathrm{2}}) \in S$. A continuous curve connecting them can be defined as
\begin{align}
    \gamma(\tau):=(U(\tau),Y(\tau)) \in S, \ \forall \ \tau \in [0,1], 
\end{align}
with boundary conditions ${\gamma(0)=(U^{\mathrm{1}},Y^{\mathrm{1}}), \gamma(1)=(U^{\mathrm{2}},Y^{\mathrm{2}})}$. The corresponding path length between them on $S$ is
\begin{align}
    \label{eq:basic_distance}
    d_S((U^{\mathrm{1}},Y^{\mathrm{1}}),(U^{\mathrm{2}},Y^{\mathrm{2}})):=\int_0^1 \norm{\gamma'(\tau)}_W d\tau.
\end{align}
An example of such a curve is
\begin{align}
    \label{eq:candidate_curve}
    \gamma(\tau)=(U^{\mathrm{1}}+\tau(U^{\mathrm{2}}-U^{\mathrm{1}}),F_{\theta}(U^{\mathrm{1}}+\tau(U^{\mathrm{2}}-U^{\mathrm{1}}))).
\end{align}
A point $(U^{\mathrm{2}},Y^{\mathrm{2}})$ can be considered to be \textit{farthest away} from $(U^{\mathrm{1}},Y^{\mathrm{1}})$ if it maximizes $d_S$. Accordingly, we formulate Problem~\eqref{eq:AL_problem} with the acquisition function
\begin{align}
\label{eq:basic_a}
    a(U|\mathcal{D}_N) = \frac{1}{N_{\mathrm{d}}} \sum_{i=1}^{N_{\mathrm{d}}}  d_S((U_i,Y_i),(U,F_{\theta}(U))),
\end{align}
such that we maximize the average path length between the new point and all the pre-existing points. Note that the summation operator can also be replaced by a $\min$ operator, resulting in a max-min formulation of Problem \eqref{eq:AL_problem}.
Unfortunately, this formulation of the acquisition function is numerically difficult to optimize, even when employing numerical integration schemes. This is because each integration step would require simulating a nonlinear dynamical system. 
In this section, we address this challenge by presenting a numerically efficient approximation of $d_S$ which exploits the qLPV nature of the underlying parameterization. We will use this approximation in the following section to develop an active learning algorithm.
\begin{remark}
To motivate the use of path length, consider the function $Y=F_{\theta}(U)$ with $U,Y \in \R$ given by $F_{\theta}(U)=0$ if $U \in [-1,0]$ and $F_{\theta}(U)=0.5\sin(2\pi U)$ if $U \in (0,1]$, and suppose $\mathcal{D}_N=\{(0,0)\}$, i.e., $N_{\mathrm{d}}=1$, with $\mathbb{U}_T=[-1,1]$. If $a(U|\mathcal{D}_N)=\sqrt{\|(0,0)-(U,F_{\theta}(U))\|_2^2}$, then Problem \eqref{eq:AL_problem} results in $U_{\mathrm{a}}=-1$, while using \eqref{eq:basic_a} results in $U_{\mathrm{a}}=1$, which is a more informative point.
\end{remark}

\begin{figure}[h]
    \centering
    \includegraphics[width=\linewidth, trim=0 0 0 0, clip]{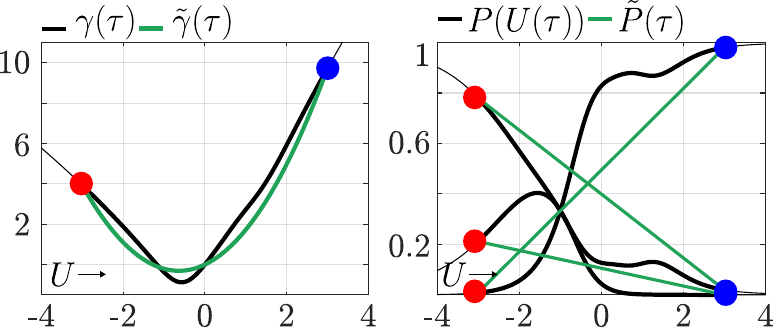}
    \caption{Illustration of the approximations in \eqref{eq:p_linear_sequence} and \eqref{eq:ignored_terms} for a random example $Y=G(P(U))U$ with $U,Y \in \R$ and $P(U) \in \R^3$. [Left] The graph $S$, with red and blue points indicating $(U^1,Y^1)$ and $(U^2,Y^2)$ respectively. The green curve is the approximation $\tilde{\gamma}(\tau)=(U(\tau),G(\tilde{P}(\tau))U)$. We report $d_S=16.6923$ and $\tilde{d}_S=16.0305$. [Right] Corresponding scheduling variable curves $P(U(\tau))$, and its approximation $\tilde{P}(\tau)$. The red dots denote $P(U^1)$ and blue $P(U^2)$.}
    \label{fig:illustration}
\end{figure}

The derivative of the curve \eqref{eq:candidate_curve} with respect to $\tau$ is
\begin{align}
    \gamma'(\tau)=\left(U^{\mathrm{2}}-U^{\mathrm{1}},\frac{dG(P(U(\tau)))U(\tau)}{d\tau}\right),
\end{align}
where the second part can be written as
\begin{align*}
    \frac{dG(P(U(\tau)))U(\tau)}{d\tau} = \Lambda(U(\tau))(U^{\mathrm{2}}-U^{\mathrm{1}}),
\end{align*}
where $\Lambda(U(\tau))$ is defined in~\eqref{eq:Lambda_defn}. Hence, we can write the path length as
\begin{align*}
\hspace{-8pt}
    d_S\scalemath{0.8}{((U^{\mathrm{1}},Y^{\mathrm{1}}),(U^{\mathrm{2}},Y^{\mathrm{2}}))}=\int_0^1 \norm{\begin{bmatrix} U^{\mathrm{2}}-U^{\mathrm{1}} \\ \Lambda(U(\tau))(U^{\mathrm{2}}-U^{\mathrm{1}}) \end{bmatrix}}_W \ d\tau.
\end{align*}
 A numerically tractable way to approximate the $d_S$ would involve considering a sequence of scalars $\{\tau_0,\ldots,\tau_M\}$ for some $M \in \mathbb{N}$, with $\tau_0=0$, $\tau_1=1$ and $\tau_{k}<\tau_{k+1}$, and approximating the path length as
\begin{align}
\label{eq:basic_distance_3}
    d_S \approxeq \sum_{k=0}^{M-1} \scalemath{0.9}{\norm{\begin{bmatrix} U(\tau_{k+1})-U(\tau_k) \\ \Lambda(U(\tau_{k+1}))U(\tau_{k+1})-\Lambda(U(\tau_{k}))U(\tau_{k}) \end{bmatrix}}}_W.
\end{align}
Note that other numerical integration schemes can be used to approximate $d_S$. While \eqref{eq:basic_distance_3}  results is a tractable formulation of the acquisition function in \eqref{eq:basic_a}, solving Problem \eqref{eq:AL_problem} with it can be numerically very expensive. This is because solving the optimization problem would involve simulating the qLPV system in \eqref{eq:qLPV} $M$-number of times to evaluate $\Lambda(U(\tau))$.
Towards reducing this complexity, we note from \eqref{eq:key_eqn} and the formulations of the regularizations in \eqref{eq:regularization_1},~\eqref{eq:regularization_2} that we explicitly aim to reduce the variation of $\Lambda(U)$ with $U$. This implies that the variation of $P(U(\tau))$ will be small along its path.
Based on this idea, we propose to approximate the scheduling variable sequence $P(U(\tau))$ as
\begin{align}
\label{eq:p_sequence}
    P(U(\tau)) \leftarrow \tilde{P}(\tau),
\end{align}
where $G(\tilde{P}(\tau))$ is a scheduling variable curve that satisfies $P(0)=P(U(0))$, $P(1)=P(U(1))$, and $G(\tilde{P}(\tau)) \in \Delta_T$ for all $\tau \in [0,1]$. 
Such a curve can be defined, for example, as
\begin{align}
\label{eq:p_linear_sequence}
   \tilde{P}(\tau)=P(U(0))+\tau(P(U(1))-P(U(0))).
\end{align}
Other options include cubic splines, Hermite polynomials, etc. While such choices might help approximate $P(U(\tau))$ better, they can be computationally more expensive to optimize.
 Using \eqref{eq:p_sequence}, we approximate \eqref{eq:basic_distance_3} as
\begin{align}
\label{eq:ignored_terms}
    \Lambda(U(\tau)) \leftarrow G(\tilde{P}(\tau)),
\end{align}
resulting in the path length approximated as
\begin{align}
\label{eq:basic_distance_4}
    \tilde{d}_S =  \sum_{k=0}^{M-1} \scalemath{0.9}{\norm{\begin{bmatrix} U(\tau_{k+1})-U(\tau_k) \\ G(\tilde{P}(\tau_{k+1}))U(\tau_{k+1})-G(\tilde{P}(\tau_{k}))U(\tau_{k}) \end{bmatrix}}}_W.
\end{align}
Finally,  we propose to use the acquisition function
\begin{align}
\label{eq:novel_a}
    a(U|\mathcal{D}_N) = \frac{1}{N_{\mathrm{d}}}  \sum_{i=1}^{N_{\mathrm{d}}}  \tilde{d}_S((U_i,Y_i),(U,F_{\theta}(U)))
\end{align}
to formulate Problem \eqref{eq:AL_problem}. This problem is computationally much lighter than when \eqref{eq:basic_a} is used, since the evaluation of $G(\tilde{P}(\tau))U$ does not require simulating the nonlinear qLPV model \eqref{eq:qLPV}, but, instead, the LTV model
\begin{align}
    \label{eq:LTV}
    x^+=A(p)x+B(p)u, && y=Cx
\end{align}
with parameter sequence $\tilde{P}(\tau)$ and input sequence $U(\tau)$. In Figure \ref{fig:illustration}, we illustrate this approximation on a random example. 
The difference between $d_S$ and $\tilde{d}_S$ depends on the difference between the curves $P(U(\tau))$ and $\tilde{P}(\tau)$, which in-turn depend on the
magnitude of the terms ignored in \eqref{eq:ignored_terms}, namely $\partial G(P(U))/\partial P_i$ and $\nabla_U P_i(U)$. Equivalently, it depends on the magnitude of the regularization coefficients, with larger values bringing them together. A theoretical analysis of this difference is a subject of future research.
\begin{remark}
    In standard settings of manifold regularization, the terms $P(U_k)$ and $P(U_l)$ in \eqref{eq:m_formulation} are replaced by $F_{\theta}(U_k)$ and $F_{\theta}(U_l)$, respectively. While this would also promote smoothness of $F_{\theta}(U)$ with respect to $U$, it does not lend itself to derive approximations similar to \eqref{eq:basic_distance_4}. One would have to directly interpolate between $Y^1$ and $Y^2$, leading to significant discrepancy between $d_S$ and $\tilde{d}_S$. In the illustrative example in Figure \ref{fig:illustration}, this would be a straight line between $(U^1,Y^1)$ and $(U^2,Y^2)$ making $\tilde{d}_S$ the Euclidean distance.
\end{remark}

\section{Active learning algorithm}

\begin{algorithm}[t]
\caption{System identification with active learning}\label{alg:al}
\begin{algorithmic}[1]
\Require $\mathcal{D}_1$, $W \succeq 0$, $M, N_{\mathrm{max}} \in \mathbb{N}$, $\epsilon_u, \kappa_1, \kappa_2 \geq 0$
\While{$N \leq N_{\mathrm{max}}$}
\State Construct $\bar{\mathbb{U}}_N$, sample $\mathcal{U}_N \subset \bar{\mathbb{U}}_N$
\State Solve Problem \eqref{eq:sysID_problem} for $\theta_N$ with either \eqref{eq:regularization_1} or \eqref{eq:regularization_2} 
\State Solve the active learning problem
\begin{align}
\label{eq:final_AL_problem}
    U_{\mathrm{a}}&=\arg\max_{U \in \bar{\mathbb{U}}_N} \ a(U|\theta_N,\mathcal{D}_N)  \\
     & \ \ \ \ \ \ \ \ \ \ \ \text{s.t.} \ \ F(U|\theta_N) \in \mathbb{Y}_T \nonumber
\end{align}
\State Run experiment on plant to collect $Y_{\mathrm{a}}=\mathbf{F}(U_{\mathrm{a}})$
\State $\mathcal{D}_{N+1} \leftarrow \mathcal{D}_N \cup \{(U_{\mathrm{a}},Y_{\mathrm{a}})\}, N \leftarrow N+1$
\EndWhile
\end{algorithmic}
\end{algorithm}

We summarize in Algorithm \ref{alg:al} the overall active learning procedure. At each step, given the current dataset $\mathcal{D}_N$ collected from the plant $Y=\mathbf{F}(U)$, we identify a model $Y=F(U|\theta_N)$ by solving Problem \eqref{eq:sysID_problem}. Then, 
we employ this model to compute a new input sequence $U_{\mathrm{a}}$ by solving Problem \eqref{eq:AL_problem}, formulated using the acquisition function in \eqref{eq:novel_a}. By apply $U_{\mathrm{a}}$ to the plant, we collect the corresponding output $Y_{\mathrm{a}}=\mathbf{F}(U_{\mathrm{a}})$ and append $(U_{\mathrm{a}},Y_{\mathrm{a}})$ to the dataset $\mathcal{D}_{N+1}$. The procedure is repeated iteratively.

The following considerations apply to the procedure:\\ \\
\indent
($i$) As per Section \ref{sec:LTV_reg}, the model $Y=F(U|\theta_N)$ can be reliably extrapolated only when $U \in \bar{\mathbb{U}}_N$. Hence, we restrict our search to this set in the active learning problem. Furthermore, even within $\bar{\mathbb{U}}_N$, the model quality might deteriorate rapidly as we move away from the previous inputs $U_i$. To penalize this deviation, we adopt the \textit{Inverse Distance Weighting} strategy presented in \citep{bemporad2023}. This is based on the normalized inverse distances
\begin{align*}
    \sigma_i(U) = \frac{1/\norm{U-U_i}_2^2}{\sum_{j=1}^{N_{\mathrm{d}}} 1/\norm{U-U_j}_2^2}&& \forall \ i \in \{1,\ldots,N_{\mathrm{d}}\},
\end{align*}
that we introduce in the acquisition function
\begin{align}
\label{eq:main_acquisition_function}
a(U|\theta_N,\mathcal{D}_N) := \sum_{i=1}^{N_{\mathrm{d}}}  \sigma_i(U) \tilde{d}_S((U_i,Y_i),(U,F(U|\theta))).
\end{align}
Note that $Y_i=\mathbf{F}(U_i)$ in \eqref{eq:main_acquisition_function}, so that we start computing the path length from the measured point.
We remark that \eqref{eq:main_acquisition_function} is equivalent to the variance function presented in \citep{bemporad2023} when formulated with $M=1$. 
\\ \\
($ii$) Since the plant $Y=\mathbf{F}(U)$ is subject to output constraints, we enforce them at the model level by imposing $F(U|\theta_N) \in \mathbb{Y}_T$ in the active learning problem. Although this approach does not guarantee that the resulting input is safe for the actual plant, the design of input filters to modify such inputs prior to experimentation is an active research topic. Nevertheless, because $F(U|\theta_N)$ is expected to model the plant well in the set $\bar{\mathbb{U}}_N$, and the active learning procedure restricts $U_{\mathrm{a}}$ to this set, empirical results indicate that the constraint $\mathbf{F}(U_{\mathrm{a}}) \in \mathbb{Y}_T$ is often satisfied.
\begin{figure*}[t]
    \centering
    \includegraphics[width=\linewidth, trim=0 0 0 0, clip]{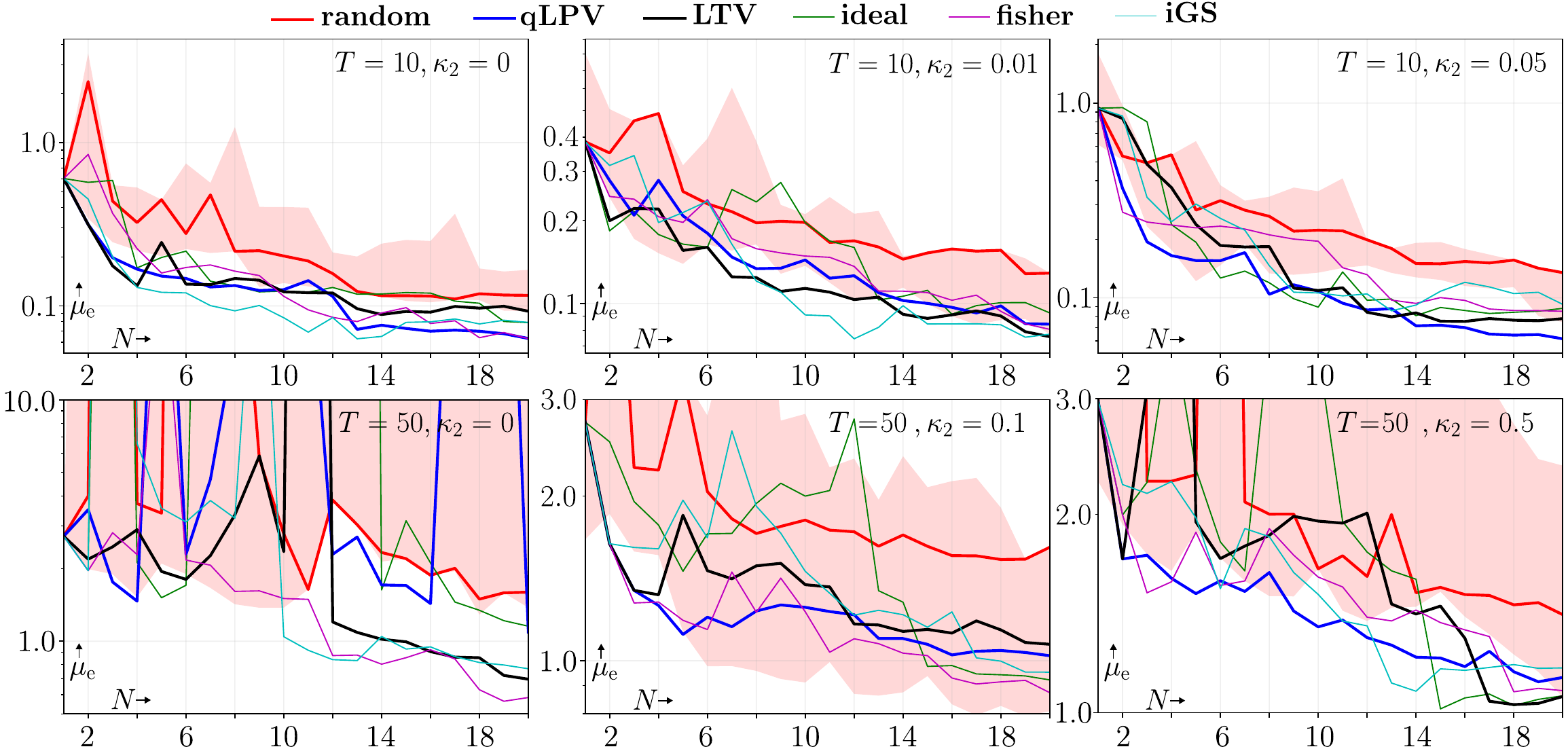}
    \caption{Results from Algorithm \ref{alg:al} - Comparison of active learning algorithm performance for $T=10$ and $T=50$. The red region is obtained by considering different seeds when using $\mathbf{random}$.}
    \label{fig:example_1}
\end{figure*}
\section{Numerical examples}
\subsection{Nonlinear oscillator}
We consider data generated by the nonlinear spring-mass-damper system with dynamics
\begin{align*}
    \mathrm{m}_{[1]}\ddot{\mathrm{x}}_{[1]}=u_{[1]}-\mathrm{k}^{\mathrm{s}}(\mathrm{x}_{[1]})-\mathrm{k}^{\mathrm{d}}(\dot{\mathrm{x}}_{[1]})-\mathrm{k}^{\mathrm{s}}(\delta \mathrm{x})-\mathrm{k}^{\mathrm{d}}(\dot{ \delta \mathrm{x}}), \\
    \mathrm{m}_{[2]}\ddot{\mathrm{x}}_{[2]}=u_{[2]}-\mathrm{k}^{\mathrm{s}}(\mathrm{x}_{[2]})-\mathrm{k}^{\mathrm{d}}(\mathrm{x}_{[2]})+\mathrm{k}^{\mathrm{s}}(\delta \mathrm{x})+\mathrm{k}^{\mathrm{d}}(\dot{ \delta \mathrm{x}}),
\end{align*}
where $(\mathrm{x}_{[1]},\mathrm{x}_{[2]})$ and $(u_{[1]},u_{[2]})$ are the positions (outputs) and forces (inputs) respectively such that $n_u=n_y=2$, $\delta \mathrm{x}=\mathrm{x}_{[1]}-\mathrm{x}_{[2]}$ is the relative position, and the spring and damper forces are $\mathrm{k}^{\mathrm{s}}(x)=\mathrm{a}x+\mathrm{b}x^3+\mathrm{c}x^5$ and $\mathrm{k}^{\mathrm{d}}(v)=\mathrm{d}v+\mathrm{e} \cdot \mathrm{tanh}(v/\mathrm{v}_\mathrm{0})$ respectively. The plant parameters are $(m_{[1]},m_{[2]},\mathrm{a},\mathrm{b},\mathrm{c},\mathrm{d},\mathrm{e},\mathrm{v}_\mathrm{0})=(2,0.01,1,10,100,1,2,0.01)$ in appropriate units. The system in simulated from the origin using the Runge-Kutta integrator (Tsit5) implemented in the $\mathrm{Diffrax}$ library \citep{diffrax} for $\mathrm{JAX}$ \citep{jax2022github}, with a time-step of $0.1$s using inputs $(u_{[1]},u_{[2]}) \in -[10,10]\times[10,10]$. We scale the inputs and outputs such that they belong inside the input and output constraint sets $\mathbb{U}$ and $\mathbb{Y}$ defined as unit boxes.

$\textit{\textbf{Model}}:$ We parameterize the qLPV system in \eqref{eq:qLPV} with $n_x=4$ and $n_p=3$, and the scheduling function in \eqref{eq:scheduling} with $h(\cdot,\cdot)=(h_1(\cdot,\cdot),\ldots,h_5(\cdot,\cdot))$, where each $h_i(\cdot,\cdot):\R^{n_x} \times \R^{n_u} \to \R$. We fix $h_1(\cdot,\cdot)=0$, and parameterize the remaining as FNNs with a single hidden layer consisting of $4$ $\mathrm{swish}$ activation units, thus satisfying Assumption \ref{ass:smooth}.

$\textit{\textbf{Experimental setup}}:$ We consider two cases with simulation lengths $T=10$ and $T=50$. We initialize each case with the dataset $\mathcal{D}_1$ built using $N_{\mathrm{d}}=5$ input-output trajectory pairs inside the constraints. We build a test set $\mathcal{D}_{\mathrm{test}}$ of $1000$ trajectory pairs, that we use to measure progress of our active learning algorithm using the  metric $\mu_{\mathrm{e}}(\theta) = \sum_{(U,Y) \in \mathcal{D}_{\mathrm{test}}}\|Y-F(U|\theta)\|_2^2/1000$ and corresponding variance $\sigma^2_{\mathrm{e}}(\theta)$ of the errors $\|Y-F(U|\theta)\|_2^2$.
We initialize the active learning algorithm with the same random parameters $\theta$ that result in $(\mu_{\mathrm{e}},\sigma^2_{\mathrm{e}})$ being $(2.0133,2.4246)$ for $T=10$ and $(19.7667,55.4161)$ for $T=50$. We choose the input neighborhood in \eqref{eq:neighborhood_U} as $\bar{\mathbb{U}}_N=\mathbb{U}_T$ with $\epsilon_u=1$, and utilize the regularization function in \eqref{eq:regularization_2} to formulate Problem \eqref{eq:sysID_problem}. At step $N$, we formulate $m(\theta)$ in \eqref{eq:m_formulation} with $N_{\mathrm{r}}=50+N_{\mathrm{d}}$ input trajectories, where the first $50$ trajectories are randomly sampled from the test set, and the remaining $N_{\mathrm{d}}$ are those in the training dataset $\mathcal{D}_N$. Over the iterations, we keep the $50$ trajectories sampled from the test set to be the same. Finally, we solve the acquisition problem in a pool-based manner by selecting an input $U_{\mathrm{a}}$ from the test set, instead of solving the problem in \eqref{eq:final_AL_problem} using an optimization algorithm. This is done to avoid issues with solver failures. Since we append these trajectories to formulate the manifold regularization function, we remove the base $50$ trajectories used in its formulation, such that the pool size is $950$ trajectories.

We formulate Problem \eqref{eq:sysID_problem} by fixing $\kappa_1=0.0001$ in \eqref{eq:regularization_2}, and choose $\kappa_2 = \{0.01,0.05\}$ for $T=10$ and $\kappa_2=\{0.1,0.5\}$ for $T=50$. We report that higher values lead to high training loss. We solve the problem by first performing $4000$ iterations of $\mathrm{Adam}$ \citep{adam} interfaced through the $\mathrm{jax}$-$\mathrm{sysid}$ toolbox \citep{Bem25}, followed by a maximum of $12000$ iterations of the $\mathrm{BFGS}$ algorithm \citep{NocedalWright2006} implemented in $\mathrm{JAX}$. We note that the best identified LTI system ($n_p=1$) with $\kappa_2=0$ and $\mathcal{D}_1=\mathcal{D}_{\mathrm{test}}$ results in $(\mu_{\mathrm{e}},\sigma^2)$ being $(0.2122,0.02803)$ for $T=10$ and $(2.4428,2.1513)$ for $T=50$ (corresponding to mean best fit rate scores \citep{Bem25} of $67.3425$ and $78.8835$ respectively), indicating a high degree of nonlinearity in the underlying plant. At each iteration, we initialize $\mathrm{Adam}$ with $\theta_{N-1}$, the solution of the previous iteration.
\begin{figure}
    \centering
    \includegraphics[width=\linewidth, trim=0 0 0 0, clip]{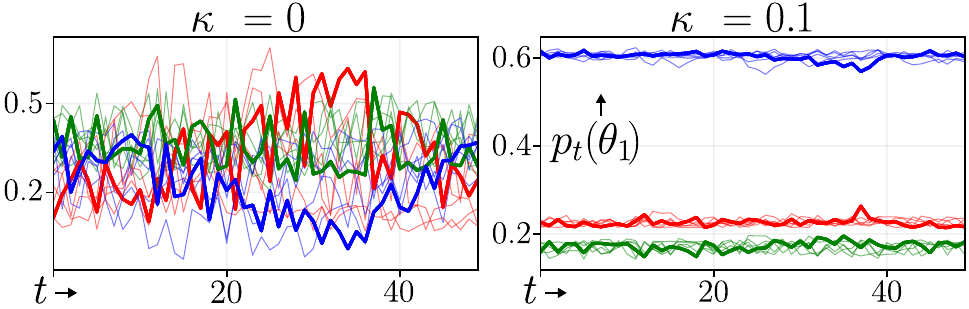}
    \caption{Scheduling variable sequences with $\mathbf{random}$ for $T=50$. The thin lines indicate $P(U|\theta_1)$ for $U$ in $\mathcal{D}_1$, and the thick line indicates $P(U_{\mathrm{a}}|\theta_1)$.}
    \label{fig:example_2}
\end{figure}
\begin{figure}
    \centering
    \includegraphics[width=\linewidth, trim=0 0 0 0, clip]{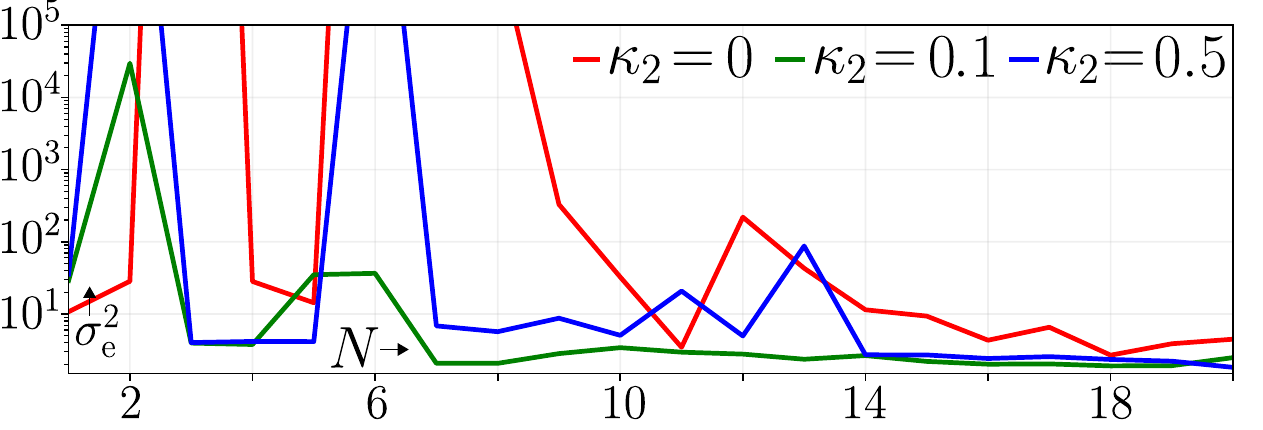}
    \caption{Error variance of $\mathbf{random}$ for $T=50$.}
    \label{fig:example_3}
\end{figure}

$\textit{\textbf{Comparisons}}:$ We formulate the functions in \eqref{eq:basic_distance_3} and \eqref{eq:basic_distance_4} with $M=10$ and $W$ the identity matrix. When the active learning objective in \eqref{eq:main_acquisition_function} is formulated with the distance function in \eqref{eq:basic_distance_3}, we label the method as $\mathbf{qLPV}$. Instead, when formulated with \eqref{eq:basic_distance_4}, as $\mathbf{LTV}$. When the objective is replaced with the variance function in \citep{bemporad2023}, we label the method as $\mathbf{ideal}$. Using classical experimental design approaches, we label the method as $\mathbf{fisher}$ \citep{goodwin1977dynamic} when using
\begin{align*}
a(U|\theta_N,\mathcal{D}_N)= \mathrm{logdet}
\scalemath{0.85}{
\begin{pmatrix*}
& \hspace{-65pt} \nabla_{\theta}F(U|\theta_N)^{\top}\nabla_{\theta}F(U|\theta_N) 
\\  & \hspace{15pt} +\sum_{i=1}^{N_{\mathrm{d}}}  \nabla_{\theta}F(U_i|\theta_N)^{\top}\nabla_{\theta}F(U_i|\theta_N)
\end{pmatrix*}}
\end{align*}
where $\nabla_{\theta}F(U|\theta_N) \in \R^{Tn_y \times n_{\theta}}$. 
Finally, when using the objective proposed in \citep[Alg. 4]{iGS}, we refer to the method as $\mathbf{iGS}$. We note that our choice of $W$ also encourages increasing the distance between input samples, which the $\mathbf{ideal}$ method does not without an explicit exploration term. The $\mathbf{iGS}$ inherently encodes such exploratory behavior. Finally, we refer to a random selection of input sequence from the pool as $\mathbf{random}$. Note that the $\textbf{random}$ approach does not consider output constraints when selecting the input $U_{\mathrm{a}}$.

$\textit{\textbf{Results}}$\footnote{Corresponding code can be found on \url{github.com/samku/qLPV_regularization}}: In Figure \ref{fig:example_1} we compare results for different values of $\kappa_2$ in \eqref{eq:regularization_2}. In general, the active learning algorithms outperform $\mathbf{random}$.  Furthermore, $\mathbf{qLPV}$ and $\mathbf{LTV}$ approaches are competitive with the other methods. We also observe that having $\kappa_2>0$ helps improve the performance of all the methods. This is attributed to the fact that manifold regularization improves extrapolation quality of the model by reducing the variance in scheduling variable trajectories over different input trajectories. In Figure \ref{fig:example_2}, we plot the scheduling variable sequences in the model $Y=F(U|\theta_1)$ when using $\mathbf{random}$ for $T=50$. The variation is much larger for $\kappa=0$ compared to $\kappa=0.1$. 
This results in reduced variance of the prediction errors over the input trajectories in the test dataset, as shown in Figure \ref{fig:example_3}. Furthermore, we report that the identified model becomes unstable on some test input trajectories at $N=3$ and $N=6$ for $\kappa_2=0$, which is avoided otherwise. The main disadvantage of using $\kappa_2>0$ is the computational complexity of Problem \eqref{eq:sysID_problem}. Future research may focus on efficiently building the sampled set $\mathcal{U}_N$ used in the formulation of \eqref{eq:m_formulation} to reduce this complexity.

\begin{figure}[h]
    \centering
    \includegraphics[width=\linewidth, trim=0 0 0 0, clip]{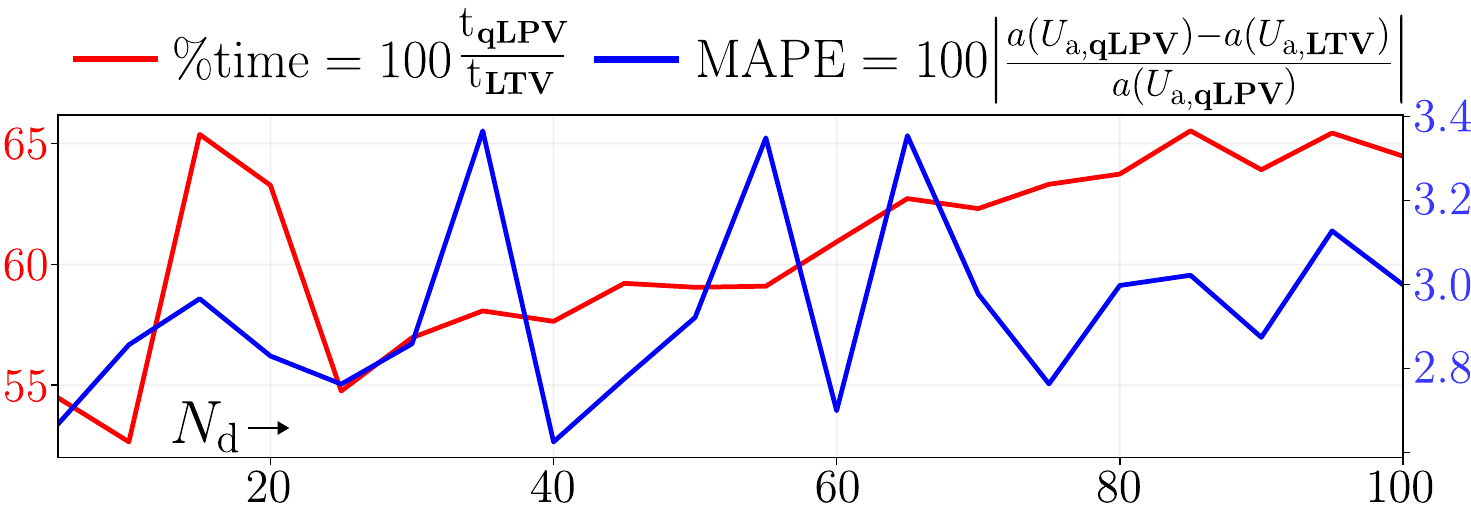}
    \caption{Solution time and suboptimality when using \eqref{eq:basic_distance_4} instead of \eqref{eq:basic_distance_3} to formulate Problem \eqref{eq:AL_problem}.}
    \label{fig:LTV_approximation}
\end{figure}

In Figure \ref{fig:LTV_approximation}, on the graph of the model $Y=F(U|\theta_{20})$ obtained using $\mathbf{qLPV}$ for $T=50$, we study the approximation error when $\tilde{d}_S$ from \eqref{eq:basic_distance_4} is used in place of $d_S$ from \eqref{eq:basic_distance_3} when formulating the cost of Problem \eqref{eq:AL_problem}, thus comparing the $\mathbf{LTV}$ and $\mathbf{qLPV}$ approaches. We fix $M=10$, randomly sample $N_{\mathrm{d}}$ number of points from the graph of the function, and solve Problem \eqref{eq:AL_problem} in a pool-based manner from a set of $1000$ randomly sampled input trajectories. In Figure \ref{fig:LTV_approximation}, we indicate $\% \mathrm{time}$ as the percentage of time required to solve the problem using \eqref{eq:basic_distance_4} as opposed to \eqref{eq:basic_distance_3}, and $\mathrm{MAPE}$ as the maximum absolute percentage error between the solutions. We observe that the $\mathbf{LTV}$ approach can solve the problem in a much quicker time with low optimality loss. We report that as $M$ increases, the $\% \mathrm{time}$ reduces further while $\mathrm{MAPE}$ remains almost the same.
\subsection{Cascaded tanks dataset}
\label{sec:cascaded}
We consider the cascaded tanks dataset presented in \citep{Cascaded_tanks}, which is measured from a fluid level control system consisting of two tanks fed by a pump. The input is the voltage of the pump that supplies water to the upper tank, and output is level of water in the lower tank. The training and test datasets are built using $T=1024$-timesteps long signals. Unlike our previous example, they do not necessarily start at the same initial state, so we need to modify the manifold regularization function in \eqref{eq:m_formulation}.
A possible approach is to modify the identification problem \eqref{eq:sysID_problem} as
\begin{align}
\label{eq:sysID_problem_init_state}
    \min_{\theta,x_0}  \|Y-F(U,x_0|\theta)\|_2^2 + r(\theta),
\end{align}
i.e., with the initial state $x_0$ as an optimization variable. Note that Problem \eqref{eq:sysID_problem_init_state} is modified for a single training input-output trajectory. The model $F(U_i,x_0|\theta)$ is obtained in the same way as in Section \ref{sec:LTV_reg}, with the modification being that the dynamics is simulated from $x_0$. Hence, a simple approach to modify the regularization function would be to redefine it as
\begin{align}
\label{eq:m_formulation_init_state}
    m(\theta)=\frac{1}{N_{\mathrm{r}}}\scalemath{0.95}{\sum_{k=1}^{N_{\mathrm{r}}-1}\sum_{l=k+1}^{N_{\mathrm{r}}} \frac{ \|P(U_k,x_0|\theta)-P(U_l,x_0|\theta)\|_2^2}{\|U_k-U_l\|_2^2},}
\end{align}
where $P(U,x_0|\theta)$ is obtained in the same way as $F(U,x_0|\theta)$. This regularization, however, would be restricted to a single initial state. To avoid this, we propose to use a multiple shooting-type approach. By fixing some $\tilde{T}<T$, let $\mathbf{I}=\{0,\tilde{T},2\tilde{T},\ldots,T-\tilde{T}\}$ and denote the input sub-trajectory in each interval as $U_{\iota}=(u_{\mathbf{I}(\iota)},\ldots,u_{\mathbf{I}(\iota)+\tilde{T}}) \in \R^{\tilde{T}n_u}$ for all $\iota \in |\mathbf{I}|$. We then sample $N_{\mathrm{r}}$ of input trajectories from the neighborhood set
$\bar{\mathbb{U}}_{\iota}=\{U_{\iota}\} \oplus \epsilon_u \mathcal{B}_{\infty}^{\tilde{T}n_u}$ for each $\iota \in \mathbf{I}$. Denoting $x_{t}$ as the state at time $t$ obtained from the model $Y=F(U,x_0|\theta)$, we utilize the manifold regularization function
\begin{align}
\label{eq:m_formulation_multi_state}
    \frac{1}{\tilde{T}N_{\mathrm{r}}}\scalemath{0.95}{\sum_{\iota=1}^{|\mathbf{I}|}\sum_{k=1}^{N_{\mathrm{r}}-1}\sum_{l=k+1}^{N_{\mathrm{r}}} \frac{ \|P(U^{\iota}_k,x_{\mathbf{I}(\iota)}|\theta)-P(U^{\iota}_l,x_{\mathbf{I}(\iota)}|\theta)\|_2^2}{\|U^{\iota}_k-U^{\iota}_l\|_2^2},}
\end{align}
where $U_k^{\iota},U_l^{\iota}$ are samples from the set $\tilde{\mathbb{U}}_{\iota}$, and the scheduling variable function is redefined to output a parameter sequence of length $\tilde{T}n_p$. 

\begin{figure}
    \centering
    \includegraphics[width=\linewidth, trim=0 0 0 0, clip]{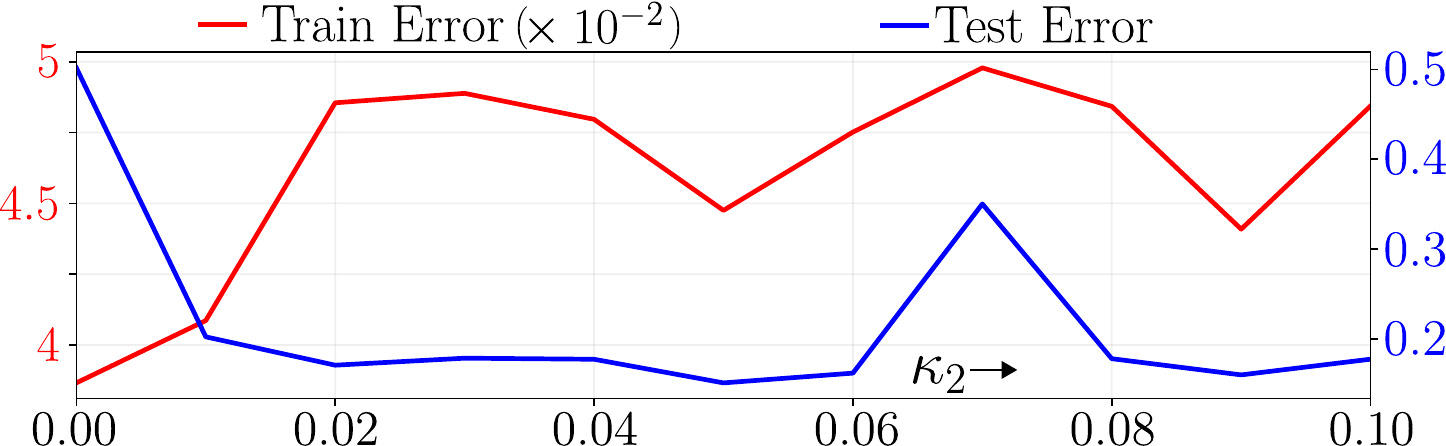}
    \caption{Error comparison for cascaded tanks benchmark.}
    \label{fig:cascaded}
\end{figure}

Utilizing \eqref{eq:m_formulation_multi_state} in lieu of $m(\theta)$, we formulate the regularization function in \eqref{eq:regularization_2} with $\epsilon_u=1$, $\tilde{T}=50$ and $N_{\mathrm{r}}=20$, that we use in Problem \eqref{eq:sysID_problem_init_state} to identify a model parameterized in the same way as in Example 1. We fix $\kappa_1=0.001$, and observe the training and test errors given the root mean square value $\sqrt{\|Y-F(U,x_0|\theta)\|_2^2/1024}.$
Over the test set, we utilize the first $5$ samples to identify the initial state. In Figure \ref{fig:cascaded}, we observe the training and test error scores for different values of $\kappa_2$. As expected, increasing $\kappa_2>0$ mildly increases the train error while being very beneficial to reduce the test error. 

\section{Conclusions}
We proposed an LTV-regularization method for qLPV system identification to promote smoothly varying dynamics, and introduced an active learning approach based on maximizing model path length along the graph of the qLPV model using this regularization for efficient optimization. Numerical examples show improved extrapolation performance and effective active learning. Future work includes handling varying initial conditions, improving sampling for constructing the regularization function, a theoretical analysis of the active learning scheme, and developing filters that ensure safe exploration.

\bibliography{ifacconf}

\begin{thebibliography}{27}
\providecommand{\natexlab}[1]{#1}
\providecommand{\url}[1]{\texttt{#1}}
\providecommand{\urlprefix}{URL }
\expandafter\ifx\csname urlstyle\endcsname\relax
  \providecommand{\doi}[1]{doi:\discretionary{}{}{}#1}\else
  \providecommand{\doi}{doi:\discretionary{}{}{}\begingroup
  \urlstyle{rm}\Url}\fi

\bibitem[{Alcalá et~al.(2019)Alcalá, Puig, and Quevedo}]{ALCALA2019}
Alcalá, E., Puig, V., and Quevedo, J. (2019).
\newblock {LPV-MPC} control for autonomous vehicles.
\newblock \emph{IFAC-PapersOnLine}, 52(28), 106--113.
\newblock 3rd IFAC Workshop on Linear Parameter Varying Systems LPVS 2019.

\bibitem[{Belkin et~al.(2006)Belkin, Niyogi, and Sindhwani}]{Belkin2006}
Belkin, M., Niyogi, P., and Sindhwani, V. (2006).
\newblock Manifold regularization: A geometric framework for learning from
  labeled and unlabeled examples.
\newblock \emph{Journal of Machine Learning Research}, 7(85), 2399--2434.

\bibitem[{Bemporad(2023)}]{bemporad2023}
Bemporad, A. (2023).
\newblock Active learning for regression by inverse distance weighting.
\newblock \emph{Information Sciences}, 626, 275--292.

\bibitem[{Bemporad(2025)}]{Bem25}
Bemporad, A. (2025).
\newblock An {L-BFGS-B} approach for linear and nonlinear system identification
  under $\ell _{1}$ and group-lasso regularization.
\newblock \emph{IEEE Transactions on Automatic Control}, 70(7), 4857--4864.

\bibitem[{Bradbury et~al.(2022)Bradbury, Frostig, Hawkins, Johnson, Leary,
  Maclaurin, Necula, Paszke, VanderPlas, Wanderman-Milne, and
  Zhang}]{jax2022github}
Bradbury, J., Frostig, R., Hawkins, P., Johnson, M.J., Leary, C., Maclaurin,
  D., Necula, G., Paszke, A., VanderPlas, J., Wanderman-Milne, S., and Zhang,
  Q. (2022).
\newblock {JAX}: Autograd and xla.

\bibitem[{Formentin et~al.(2019)Formentin, Mazzoleni, Scandella, and
  Previdi}]{Formentin2019}
Formentin, S., Mazzoleni, M., Scandella, M., and Previdi, F. (2019).
\newblock Nonlinear system identification via data augmentation.
\newblock \emph{Systems \& Control Letters}, 128, 56--63.

\bibitem[{Gevers et~al.(2011)Gevers, Bombois, and Hildebrand}]{Gevers2011}
Gevers, M., Bombois, X., and Hildebrand, R. (2011).
\newblock Optimal experiment design for open and closed-loop system
  identification.
\newblock \emph{Communications in information and systems}, 11, 197--224.

\bibitem[{Goodwin and Payne(1977)}]{goodwin1977dynamic}
Goodwin, G.C. and Payne, R.L. (1977).
\newblock \emph{Dynamic System Identification: Experiment Design and Data
  Analysis}, volume 136 of \emph{Mathematics in Science and Engineering}.
\newblock New York.

\bibitem[{He(2010)}]{He2010}
He, X. (2010).
\newblock Laplacian regularized d-optimal design for active learning and its
  application to image retrieval.
\newblock \emph{IEEE Transactions on Image Processing}, 19(1), 254--263.

\bibitem[{Kidger(2021)}]{diffrax}
Kidger, P. (2021).
\newblock \emph{{O}n {N}eural {D}ifferential {E}quations}.
\newblock Ph.D. thesis, University of Oxford.

\bibitem[{Kingma and Ba(2015)}]{adam}
Kingma, D.P. and Ba, J. (2015).
\newblock Adam: A method for stochastic optimization.
\newblock In \emph{Proceedings of the 3rd International Conference on Learning
  Representations (ICLR)}.
\newblock The paper was initially released on arXiv in 2014.

\bibitem[{Mania et~al.(2022)Mania, Jordan, and Recht}]{mania2022active}
Mania, H., Jordan, M.I., and Recht, B. (2022).
\newblock Active learning for nonlinear system identification with guarantees.
\newblock \emph{Journal of Machine Learning Research}, 23(32), 1--30.

\bibitem[{Mulagaleti and Bemporad(2024)}]{mulagaleti2024a}
Mulagaleti, S.K. and Bemporad, A. (2024).
\newblock Combined learning of linear parameter-varying models and robust
  control invariant sets.

\bibitem[{Mulagaleti and Bemporad(2025)}]{mulagaleti2025}
Mulagaleti, S.K. and Bemporad, A. (2025).
\newblock Learning quasi-{LPV} models and robust control invariant sets with
  reduced conservativeness.
\newblock \emph{IEEE Control Systems Letters}, 9, 252--257.

\bibitem[{Nocedal and Wright(2006)}]{NocedalWright2006}
Nocedal, J. and Wright, S.J. (2006).
\newblock \emph{Numerical Optimization}.
\newblock New York, NY, 2 edition.

\bibitem[{Ohlsson et~al.(2008)Ohlsson, Roll, and Ljung}]{Ohlsson1008}
Ohlsson, H., Roll, J., and Ljung, L. (2008).
\newblock Manifold-constrained regressors in system identification.
\newblock In \emph{2008 47th IEEE Conference on Decision and Control},
  1364--1369.

\bibitem[{Piga et~al.(2024)Piga, Rufolo, Maroni, Mejari, and
  Forgione}]{Piga2024}
Piga, D., Rufolo, M., Maroni, G., Mejari, M., and Forgione, M. (2024).
\newblock Synthetic data generation for system identification: leveraging
  knowledge transfer from similar systems.
\newblock In \emph{2024 IEEE 63rd Conference on Decision and Control (CDC)},
  6383--6388.

\bibitem[{Rojas et~al.(2007)Rojas, Welsh, Goodwin, and Feuer}]{Rojas2007}
Rojas, C.R., Welsh, J.S., Goodwin, G.C., and Feuer, A. (2007).
\newblock Robust optimal experiment design for system identification.
\newblock \emph{Automatica}, 43(6), 993--1008.

\bibitem[{Schoukens and Noël(2017)}]{Cascaded_tanks}
Schoukens, M. and Noël, J. (2017).
\newblock Three benchmarks addressing open challenges in nonlinear system
  identification.
\newblock \emph{IFAC-PapersOnLine}, 50(1), 446--451.
\newblock 20th IFAC World Congress.

\bibitem[{Settles(2009)}]{settles2009active}
Settles, B. (2009).
\newblock Active learning literature survey.
\newblock Technical Report 1648, University of Wisconsin-Madison.

\bibitem[{T{\'o}th(2010)}]{Toth2010Modeling}
T{\'o}th, R. (2010).
\newblock \emph{Modeling and Identification of Linear Parameter-Varying
  Systems}, volume 403 of \emph{Lecture Notes in Control and Information
  Sciences}.
\newblock Berlin, Heidelberg.

\bibitem[{Verhoek et~al.(2022)Verhoek, Beintema, Haesaert, Schoukens, and
  Tóth}]{Verhoek2022}
Verhoek, C., Beintema, G.I., Haesaert, S., Schoukens, M., and Tóth, R. (2022).
\newblock Deep-learning-based identification of {LPV} models for nonlinear
  systems.
\newblock In \emph{2022 IEEE 61st Conference on Decision and Control (CDC)},
  3274--3280.

\bibitem[{Wagenmaker and Jamieson(2020)}]{wagenmaker20a}
Wagenmaker, A. and Jamieson, K. (2020).
\newblock Active learning for identification of linear dynamical systems.
\newblock In J.~Abernethy and S.~Agarwal (eds.), \emph{Proceedings of Thirty
  Third Conference on Learning Theory}, volume 125 of \emph{Proceedings of
  Machine Learning Research}, 3487--3582.

\bibitem[{Wagenmaker et~al.(2023)Wagenmaker, Shi, and
  Jamieson}]{wagenmaker2023optimal}
Wagenmaker, A., Shi, G., and Jamieson, K.G. (2023).
\newblock Optimal exploration for model-based {RL} in nonlinear systems.
\newblock \emph{Advances in Neural Information Processing Systems}, 36,
  15406--15455.

\bibitem[{Wu et~al.(2019)Wu, Lin, and Huang}]{iGS}
Wu, D., Lin, C.T., and Huang, J. (2019).
\newblock Active learning for regression using greedy sampling.
\newblock \emph{Information Sciences}, 474, 90--105.

\bibitem[{Zhang et~al.(2016)Zhang, Shum, and Shao}]{zhang2016manifold}
Zhang, L., Shum, H.P., and Shao, L. (2016).
\newblock Manifold regularized experimental design for active learning.
\newblock \emph{IEEE Transactions on Image Processing}, 26(2), 969--981.

\bibitem[{Zhou et~al.(2003)Zhou, Bousquet, Lal, Weston, and
  Sch\"{o}lkopf}]{Zhou2003}
Zhou, D., Bousquet, O., Lal, T., Weston, J., and Sch\"{o}lkopf, B. (2003).
\newblock Learning with local and global consistency.
\newblock In S.~Thrun, L.~Saul, and B.~Sch\"{o}lkopf (eds.), \emph{Advances in
  Neural Information Processing Systems}, volume~16.

\end{thebibliography}

\end{document}